\documentclass[12pt]{article}
\usepackage{graphicx}

\usepackage{hyperref}   

\newcommand\pubnumber{CMS CR-2019/262}  
\newcommand\pubdate{\today}

\def\institute{Department of Physics and Astronomy\\
University of Bologna and INFN, I-40127 Bologna, ITALY}

\def\Title#1{\begin{center} {\Large #1 } \end{center}}
\def\Author#1{\begin{center}{ \sc #1} \end{center}}
\def\Address#1{\begin{center}{ \it #1} \end{center}}

\newcommand\pubblock{\rightline{\begin{tabular}{l} \pubnumber\\
         \pubdate  \end{tabular}}}
\newenvironment{Abstract}{\begin{quotation}  }{\end{quotation}}
\newenvironment{Presented}{\begin{quotation} \begin{center} 
             PRESENTED AT\end{center}\bigskip 
      \begin{center}\begin{large}}{\end{large}\end{center} \end{quotation}}

\def\beq{\begin{equation}}
\def\eeq#1{\label{#1}\end{equation}}
\def\eeqn{\end{equation}}

\def\beqa{\begin{eqnarray}}
\def\eeqa#1{\label{#1}\end{eqnarray}}
\def\eeqan{\end{eqnarray}}

\let\bar=\overbar

\def\Dslash{\not{\hbox{\kern-4pt $D$}}}
\def\dslash{\not{\hbox{\kern-2pt $\del$}}}

\def\msb{{\bar{\ssstyle M \kern -1pt S}}}

\begin{document}
\begin{titlepage}
\pubblock

\vfill

\Title{Top Quark Mass Measurements in ATLAS and CMS}

\vfill

\Author{\textsc{Andrea Castro, on behalf of the ATLAS and CMS Collaborations}}
\Address{\institute}
\vfill

\begin{Abstract}
The ATLAS and CMS Collaborations have performed a variety of measurements of the top quark mass, taking advantage of the abundant production of top quarks at the LHC.
 The most recent measurements are reported here, based on data collected at 8 and 13 TeV, with particular emphasis on the distinction between the so-called ``direct'' measurements and the ``indirect'' evaluations obtained from cross sections and differential cross sections.
\end{Abstract}
\vfill
\begin{Presented}
$12^{th}$ International Workshop on Top Quark Physics\\
Beijing, China,  September 22--27, 2019
\end{Presented}
{\small
\bigskip
\bigskip
\bigskip
\textcopyright 2019 CERN for the benefit of the ATLAS and CMS Collaborations.
\par\noindent
Reproduction of this article or parts of it is allowed as specified in the CC-BY-4.0 license.
}
\vfill
\end{titlepage}
\def\thefootnote{\fnsymbol{footnote}}
\setcounter{footnote}{0}

\section{Introduction}
 At the CERN LHC, top quarks are produced in proton-proton ($pp$) collisions  mainly as top quark-antiquark ($t\bar t$) pairs. This production is quite abundant: more than 5 million $t\bar t$ pairs have been produced in the Run 1 at $\sqrt{s}=7$ and 8 TeV, and more than 100 million in the Run 2 at $\sqrt{s}=13$ TeV. The events collected from the chain  $pp\to t\bar t\to W^+bW^-\bar b$ have a  final state distinguished by the decay of the two $W$ bosons into {\it dilepton}, {\it lepton+jets}, and {\it 
all-jets} channels. 
\par
The top quark mass ($m_t$) is a free parameter of the standard model (SM), which can be measured ``directly'' from the reconstruction of the decay products of the top quark, or 
``indirectly'' comparing the $t\bar t$ cross section (inclusive or differential) to theoretical expectations, 
yielding a value determined in a well-defined theoretical scheme, e.g. the pole mass or the $\bar{\mathrm{MS}}$ mass, as discussed for instance in~\cite{nason}. 
\par
Precise measurements of $m_t$, the $W$ boson mass, and the Higgs boson mass are crucial, and can be used to test the self-consistency of the SM~\cite{sm-global-fit}, to explore models for new physics~\cite{bsm-ref}, or to assess the vacuum stability~\cite{vacuum-stability} of the SM.
\par
 In the following, recent analyses carried by the ATLAS and CMS Collaborations are reported, mentioning the values of $\sqrt{s}$ and integrated luminosity.
\section{Direct measurement of the top quark mass}
The $t\bar t$  events collected by  ATLAS~\cite{atlas-ref} and CMS~\cite{cms-ref} in $pp$ collisions have common physics 
signatures: isolated leptons ($e$ or $\mu$) with high transverse momentum ($p_{\rm T}$); high-$p_{\rm T}$ jets, some of which associated to the hadronization of $b$ quarks ($b$-jets, i.e. $b$-tagged); missing transverse momentum ($p_{\rm T}^{\rm miss}$)  associated to neutrinos.
 In addition, because of the large  value of $\sqrt{s}$, top quarks can be produced at  high $p_{\rm T}$, originating the so-called  {\em boosted jets}, where the decay products of the top quark merge into a single wide jet. Using these physics objects, the $t\bar t\to WbWb$ final state can be reconstructed, and kinematic distributions of variables sensitive to $m_t$ can be obtained.
\par
{\bf ATLAS: lepton+jets events at 8 TeV (20.2 fb$^{-1}$).}
Events are selected~\cite{atlas-ljets-ref}  requiring the presence of one electron or muon, at least four jets, two of which $b$-tagged.
 A selection based on a boosted decision tree is applied to improve the correct kinematical matching to the $t\bar t\to WbWb$ final state. A {\em template method} is applied, using in addition to the invariant mass of the reconstructed top quark and of the W boson, also the ratio between the $p_T$ sum of tagged and untagged jets, $R_{bq}^{reco}=(p_T^{b1}+p_T^{b2})/(p_T^{j1}+p_T^{j2})$, for the in situ calibration of the $b$-jet energy scale factors.  
The fit returns a value for $m_t$ of
 $172.08\pm 0.39\,({\rm stat})\, \pm 0.82\,({\rm syst})$ GeV, with a total relative uncertainty of $0.53\%$.
 The systematic uncertainty is dominated by  the  jet energy scale (JES) uncertainty (0.54 GeV), the $b$-tagging modeling  (0.38 GeV), and the jet energy resolution (0.20 GeV). A combination~\cite{atlas-ljets-ref} of measurements at 7 and 8 TeV in the lepton+jets, dilepton and all-jets channels yields a top quark mass of   $172.69\pm 0.48$ GeV, with a relative uncertainty of $0.28\%$.
\par
{\bf CMS: dilepton events at 13 TeV (35.9 fb$^{-1}$).}
In this case, the value of the top quark mass comes from a simultaneous fit of $m_t$ and of the production cross section~\cite{cms-dilepton-ref}. Events are selected requiring one electron and one muon with opposite charge, and defining categories based on the number of $b$-jets and of additional jets. The minimum invariant mass of the lepton+$b$-jet system is used in a template method, yielding $m_t=172.33\pm 0.14\,({\rm stat})\, ^{+0.66}_{-0.72}\,({\rm syst})$ GeV, with a total relative uncertainty of approximately $0.42\%$, mainly due to the JES uncertainty (0.57 GeV), the limited size of Monte Carlo samples (0.36 GeV), and the background modeling  (0.24 GeV).
\par
{\bf CMS: all-jets events at 13 TeV (35.9 fb$^{-1}$).}
Events with at least six jets, two of which $b$-tagged, are selected for this analysis. The measurement~\cite{cms-alljets-ref} is based on the so-called {\em ideogram method} where the  kinematical reconstruction of the  $WbWb$ final state is followed by a likelihood fit, based on distributions of the $W$ boson and top quark reconstructed masses, allowing for an in situ calibration of the jet scale factor (JSF). The value of top quark mass returned by the fit is $m_t=172.34\pm 0.20\,({\rm stat}+{\rm JSF})\pm 0.43\,({\rm CR+ERD})\pm 0.55\,({\rm syst})$ GeV, with a total uncertainty of 0.73 GeV ($0.42\%$).   The systematic uncertainty is mainly due to an advanced treatment of the modeling of color reconnection (CR) and early resonance decays (ERD), and to flavor-related jet energy corrections (0.34 GeV). A combination with results from the lepton+jets channel from the same data set yields $m_t=172.26\pm 0.61$ GeV, with a relative uncertainty of $0.36\%$.
\par
{\bf ATLAS: lepton+jets with an additional soft $\mu$ at 13 TeV (36.1 fb$^{-1}$).}
This new result~\cite{atlas-ljets-mu-ref}, presented for the first time here, is based on events selected requiring one electron or muon, at least four jets, one of which $b$-tagged by a displaced vertex algorithm, and another one with a soft muon inside.   The invariant mass of the primary lepton and the soft muon is used in a template method, distinguishing same-sign and opposite-sign events. The fit yields $m_t=174.48\pm 0.40\,({\rm stat})\, \pm 0.67\,({\rm syst})$ GeV, with a total relative uncertainty of  $0.45\%$. Major contributions to the systematic uncertainty come from the heavy-flavor decay (0.39 GeV) and pileup (0.20 GeV) 
modeling.

\section{Indirect determination of the top quark mass}
An alternative approach is to extract the top quark mass ``indirectly'', comparing measured inclusive and differential cross sections to theoretical calculations as a function of an assumed pole mass ($m_t^{pole}$) or  $\bar{\mathrm{MS}}$  mass ($m_t(m_t)$). 
\par
\begin{figure}[htb]
\begin{tabular}{ccc}
\begin{minipage}{0.4\textwidth}
\includegraphics[width=6.5cm]{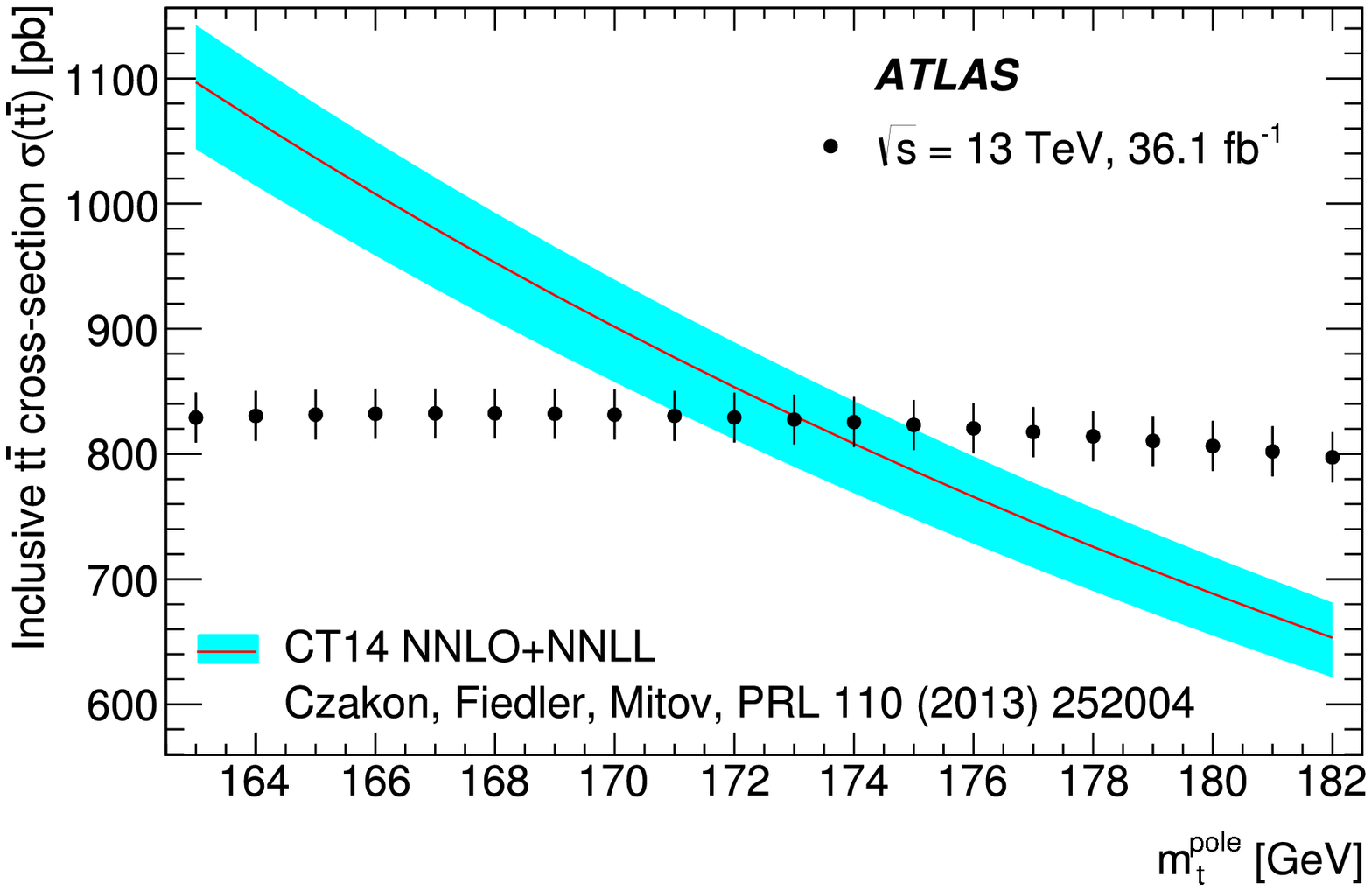}
\end{minipage}
& ~~~~~ & 
\begin{minipage}{0.4\textwidth} 
\includegraphics[width=6.5cm]{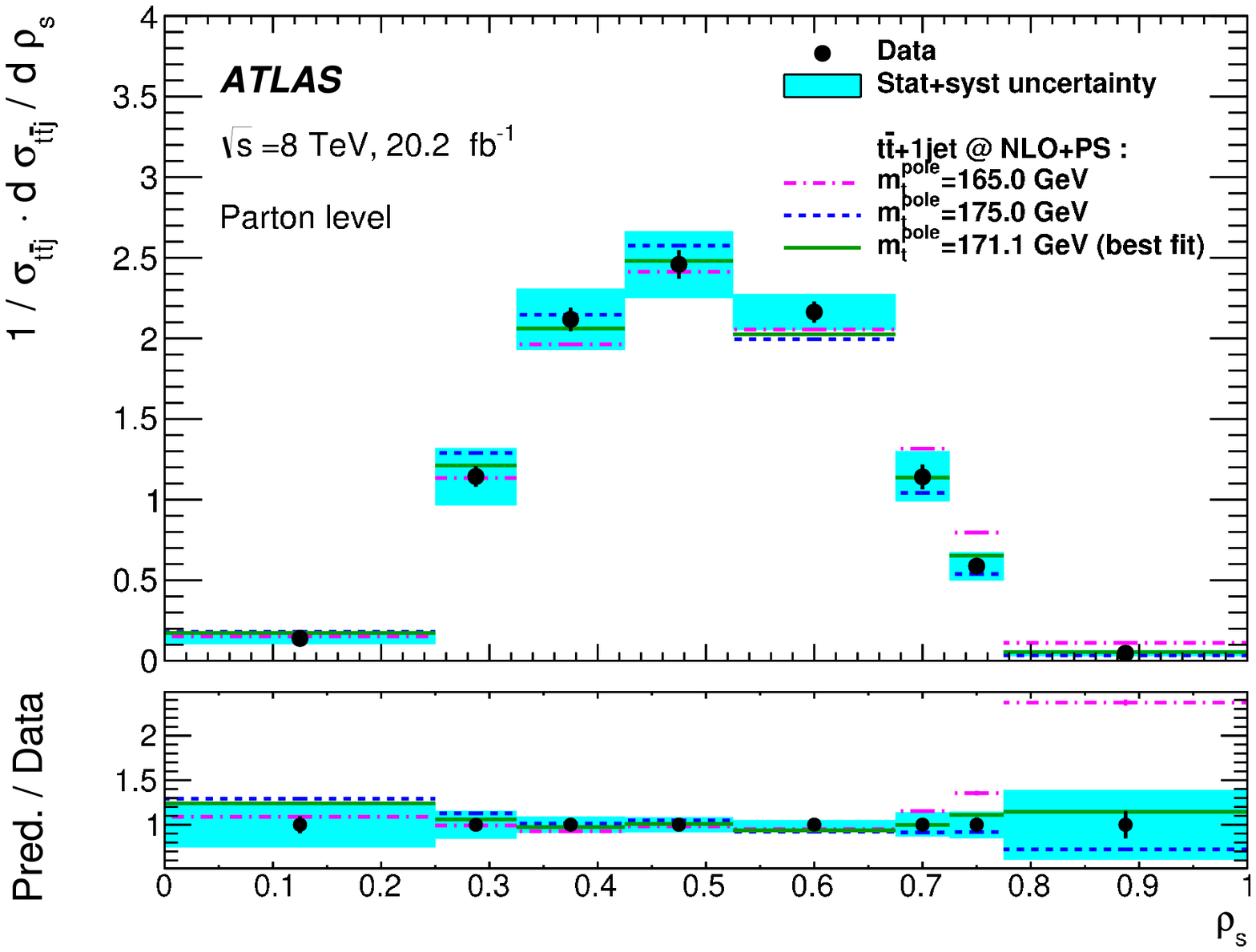}
\end{minipage}
\end{tabular}
\caption{ATLAS results. Left: inclusive $t\bar t$ cross section $vs$ $m_t^{pole}$~\cite{atlas-xsecmt-ref}. Right: normalized differential cross section as a function of $\rho_s$~\cite{atlas-tt1jet-ref}.}
\label{mass-atlas}
\end{figure}

{\bf ATLAS: inclusive $t\bar t$ cross section in the $e\mu$ channel at  13 TeV (36.1 fb$^{-1}$).}
The inclusive $t\bar t$ production cross section is measured~\cite{atlas-xsecmt-ref} from opposite-charged $e\mu$ events with one or two $b$-jets. The cross section is expressed as a function of $m_t^{pole}$, and compared to a prediction at next-to-next-to-leading order with next-to-next-to-leading-log resummation, see Fig.\,\ref{mass-atlas}-left. The resulting value is $m_t^{pole}=173.1\, ^{+2.0}_{-2.1}$ GeV, where the major contributions to the systematic uncertainties are associated to the choice of parton distribution functions (PDFs) and $\alpha_S$, and variations in QCD scales.
\par
{\bf ATLAS: differential cross section for lepton+jets $t\bar t$+1jet events at 8~TeV (20.2 fb$^{-1}$).}
Events are selected  requiring one electron or muon and at least five jets, two of which $b$-tagged. The normalized differential cross section is derived~\cite{atlas-tt1jet-ref} as a function of the variable 
$\rho_s=2m_0/m_{t\bar tj}$, where $m_0=170$ GeV is a reference mass, while $m_{t\bar tj}$ is the invariant mass of the $t\bar t+1$jet system.
A comparison of this differential cross section to next-to-leading-order predictions, see Fig.\,\ref{mass-atlas}-right,  yields a value 
$m_t^{pole}=171.1\pm 0.4\,({\rm stat})\,\pm 0.9\,({\rm syst})\, ^{+0.7}_{-0.3}\,({\rm theo})$ GeV,
 with systematic uncertainties  dominated by experimental contributions. 
A similar procedure is applied to derive the $\bar{\mathrm{MS}}$ mass $m_t(m_t)=162.9\pm 0.5\,({\rm stat})\,\pm 1.0\,({\rm syst})\, ^{+2.1}_{-1.2}\,({\rm theo})$ GeV, with larger systematic uncertainties associated to scale variations.
\par
\begin{figure}[htb]
\begin{tabular}{ccc}
\begin{minipage}{0.4\textwidth}
\includegraphics[width=6.5cm]{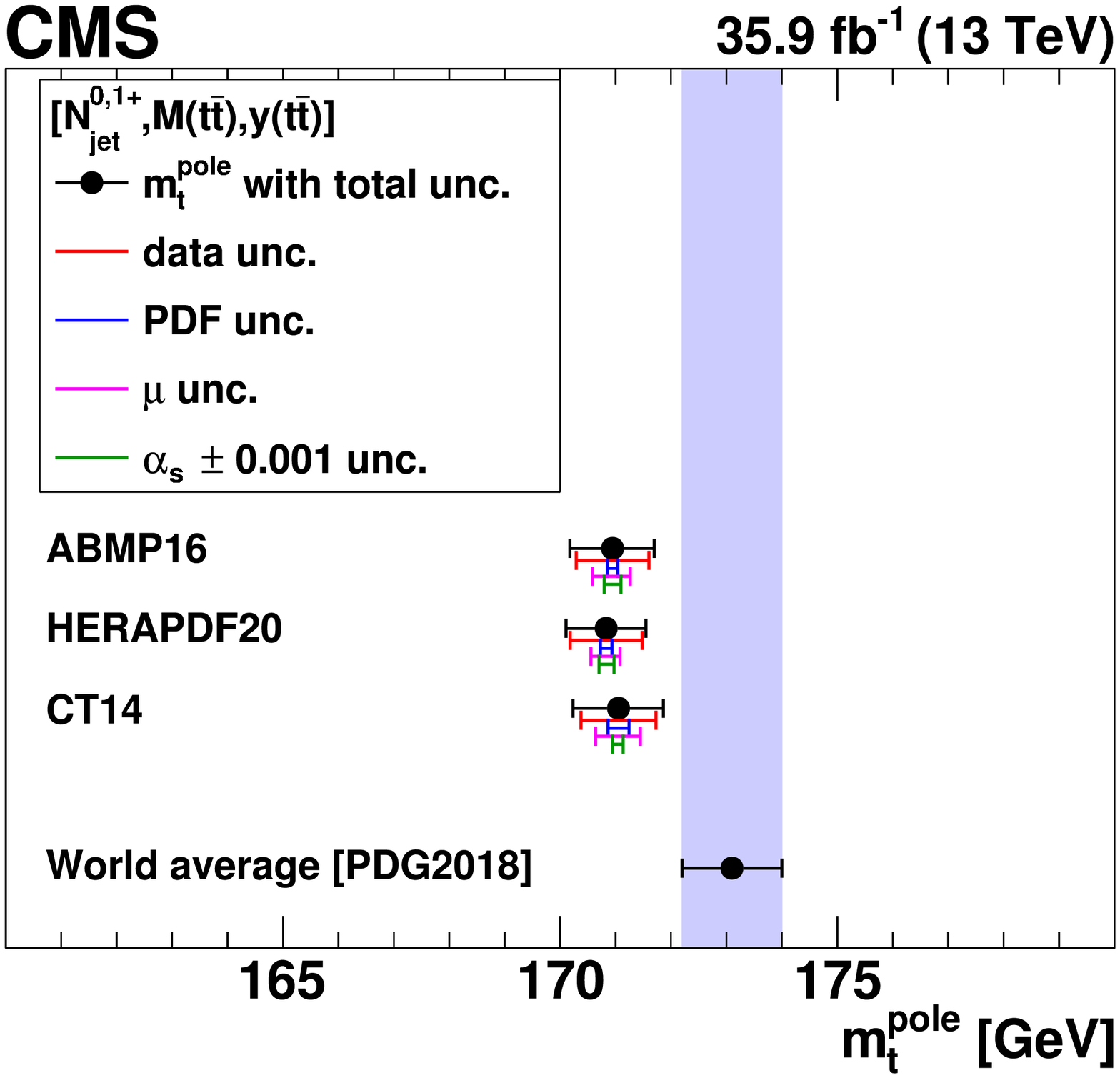}
\end{minipage}
& ~~~~~ &
\begin{minipage}{0.4\textwidth} 
\includegraphics[width=7.0cm]{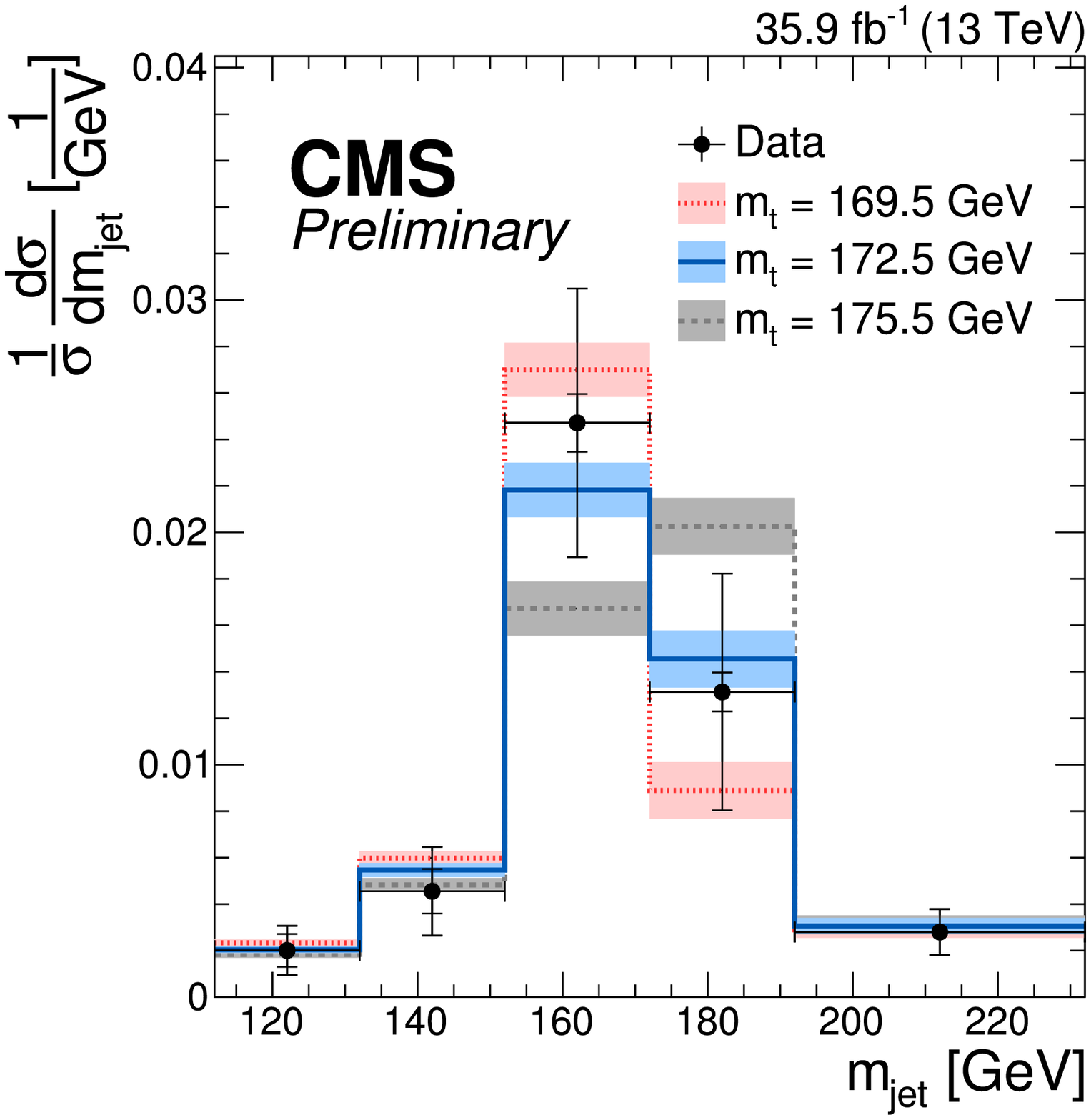}
\end{minipage}
\end{tabular}
\caption{CMS results. Left: $m_t^{pole}$ values extracted at NLO using different PDF sets~\cite{cms-3xsec-ref}. Right: 
normalized differential cross section as a function of $m_{jet}$~\cite{cms-mjet-ref}.}
\label{mass-cms}
\end{figure}

{\bf CMS: triple-differential cross section in dilepton events at 13 TeV 
\par\noindent (35.9~fb$^{-1}$).}
The event selection  requires 2 electrons or muons of opposite charge, and at least two jets, one of which $b$-tagged. A normalized triple-differential cross section is derived~\cite{cms-3xsec-ref} as a function of the invariant mass and rapidity of the $t\bar t$ system, and of the number of  additional jets not coming from the top quark decay. From a comparison to fixed-order next-to-leading-order predictions, $m_t^{pole}$ and $\alpha_S$ are extracted, yielding $m_t^{pole}=170.5\pm 0.8$ GeV, see Fig.\,\ref{mass-cms}-left, with uncertainties dominated by the fit procedure (0.7 GeV) and the QCD scales (0.3 GeV). A precision smaller than 1 GeV has been reached, similar to what obtained for direct measurements.
\par
{\bf CMS: invariant jet mass distribution for boosted jets in lepton+jets events at 13 TeV (35.9 fb$^{-1}$).}
The location of the peak in the jet invariant mass ($m_{jet}$) distribution is sensitive to $m_t$, as calculated in soft collinear effective theory~\cite{sceft}. 
In this case~\cite{cms-mjet-ref}, events are selected  requiring one electron or muon, at least one $b$-jet and at least one boosted wide jet. 
The wide jet is reconstructed with XCone~\cite{xcone-ref}, a novel algorithm suitable to reconstruct wide jets and the subjets inside them, 
 improving the $m_{jet}$ resolution and reducing underlying event and pileup effects. 
The normalized differential cross section as a function of $m_{jet}$ is compared in Fig.\,\ref{mass-cms}-right to theoretical predictions giving $m_t=172.56\pm 2.47$ GeV, with systematic uncertainties dominated by experimental effects (1.58 GeV) and signal modeling (1.55 GeV).
\par
{\bf CMS: running of the top quark mass in $e\mu$ events at  13 TeV (35.9 fb$^{-1}$).}
The differential cross section as a function of the mass of the $t\bar t$ system is extracted at parton level from multidifferential distributions as a function of the 
$t\bar t$ mass, the smallest invariant mass of the lepton+$b$-jet system, and the $p_T$ of the softest jet~\cite{cms-mtrun-ref}.
 Values of  $\bar{\mathrm{MS}}$ mass are extracted in each one of the bins of the differential cross section, at a scale $\mu_k$ given by the average mass in each bin $k$. The observed evolution of the $m_t(\mu_k)$ values agrees with the prediction from renormalization group equations at 1-loop precision within 1.1 standard deviations. The no-running hypothesis is excluded at 95\% confidence level.

\section{Summary}
Since the  top quark discovery, made 24 years ago, the ``direct'' measurement of its mass has been actively pursued at $p\bar p$/$pp$ colliders recurring to a variety of channels and techniques.
The precision reached is quite impressive, smaller than $0.3\%$, and expected to improve with ongoing and future measurements at the LHC, including refinements in the methodology.  New attention is given to ``indirect'' measurements of the top quark mass through comparisons of inclusive and differential cross sections to predictions as a function of better theoretically-defined parameters as the pole mass or the $\bar{\mathrm{MS}}$ mass. In the future, it will be more and more important to reduce the systematic uncertainties, mainly those related to signal modeling, with a thorough  tuning of the parameters in the Monte Carlo generators, and also reduce the theoretical ambiguities in the definition of the top quark mass itself.


\begin{thebibliography}{99}

\bibitem{nason} P. Nason, arXiv:1712.02796.

\bibitem{sm-global-fit} M. Baak et al.,
Eur. Phys. J. C {\bf 74} (2014) 3046.

\bibitem{bsm-ref} C. T. Hill, Phys. Lett. B {\bf 266} (1991) 419; C. T. Hill,
Phys. Lett. B {\bf 345} (1995) 483; W. A. Bardeen, C. T. Hill and M. Lindner,
Phys. Rev. D {\bf 41} (1990) 1647.

\bibitem{vacuum-stability} D. Buttazzo et al.,
JHEP {\bf 12} (2013) 089.


\bibitem{atlas-ref} ATLAS Collaboration,  JINST {\bf 3} (2008) S08003.

\bibitem{cms-ref} CMS Collaboration,  JINST {\bf 3} (2008) S08004.


\bibitem{atlas-ljets-ref} ATLAS Collaboration, 
Eur. Phys. J. C {\bf 79} (2019) 290; arXiv:1810.01772.

\bibitem{cms-dilepton-ref} CMS Collaboration, 
Eur. Phys. J. C {\bf 79} (2019) 368; arXiv:1812.10505.

\bibitem{cms-alljets-ref} CMS Collaboration, 
Eur. Phys. J. C {\bf 79} (2019) 313; arXiv:1812.10534.


\bibitem{atlas-ljets-mu-ref} ATLAS Collaboration, 
 \href{https://cds.cern.ch/record/2693954}{ATLAS-CONF-2019-046}. See also the poster presented by L. Wilkins, this conference.

\bibitem{atlas-xsecmt-ref} ATLAS Collaboration, 
 \href{https://cds.cern.ch/record/2686255}{ATLAS-CONF-2019-041}.

\bibitem{atlas-tt1jet-ref} ATLAS Collaboration, 
arXiv:1905.02302, submitted to JHEP.

\bibitem{cms-3xsec-ref} CMS Collaboration, 
arXiv:1904.05237, submitted to Eur. Phys. J. C.

\bibitem{sceft} A.H. Hoang, S. Pl\"atzer and D. Samitz,
JHEP 10 (2018) 200.

\bibitem{cms-mjet-ref} CMS Collaboration, 
CMS Physics Analysis Summary, \href{https://cds.cern.ch/record/2682624}{ CMS-PAS-TOP-19-005}.

\bibitem{xcone-ref} I.W. Stewart, F.J. Tackmann, J. Thaler et al.,  
JHEP 11 (2015) 072.

\bibitem{cms-mtrun-ref} CMS Collaboration, 
arXiv:1909.09193, submitted to Phys. Lett. B. See also the presentation given by M. Defranchis, this conference.


\end{thebibliography}
\end{document}